\documentclass[useAMS,usenatbib]{mn2e}
\usepackage{epsfig}
\usepackage{natbib}

\title{Spectral Energy Distribution variation in BL Lacs and FSRQs}
\author[Rani et al.]
{Bindu Rani$^{1,2,3}$\thanks{E-mail: bindu@aries.res.in},
Alok C.\ Gupta$^{1,2}$,
R. Bachev$^{3}$,
A. Strigachev$^{3}$,
E. Semkov$^{3}$,
\newauthor   F. D'Ammando$^{4}$,
P. J. Wiita$^{5}$, 
M. A. Gurwell$^{6}$,
E. Ovcharov$^{7}$,
B. Mihov$^{3}$,
\newauthor  S. Boeva$^{3}$,
S. Peneva$^{3}$   \\
$^{1}$Aryabhatta Research Institute of Observational Sciences (ARIES),
Manora Peak, Nainital -- 263129, India\\
$^{2}$Department of Physics, DDU Gorakhpur University, Gorakhpur-273009, India  \\
$^{3}$Institute of Astronomy and National Astronomical Observatory,                
Bulgarian Academy of Sciences, 72 Tsarigradsko Shosse Blvd., 1784 Sofia, Bulgaria  \\
$^{4}$INAF-IASF Palermo, Via Ugo La Malfa 153, I-90146 Palermo, Italy \\
$^{5}$Department of Physics, The College of New Jersey, P.O.\ Box 7718, Ewing, NJ 08628, USA  \\
$^{6}$Harvard-Smithsonian Center for Astrophysics, Cambridge, MA 02138 USA  \\
$^{7}$Department of Astronomy, University of Sofia, 5 James Bourchier, 1164 Sofia, Bulgaria
}

\begin{document}

\date{Accepted ....... Received  ......; in original form ......}

\pagerange{\pageref{firstpage}--\pageref{lastpage}} \pubyear{2011}

\maketitle

\label{firstpage}

\begin{abstract}
 We present the results of our study of spectral energy distributions (SEDs) of a sample 
of ten low- to intermediate-synchrotron-peaked blazars.  We investigate some of the physical parameters 
most likely responsible for the observed short-term variations in blazars.  To do so, we 
focus on the study of changes in the SEDs of blazars corresponding to changes in their 
respective optical fluxes. We model the observed spectra of blazars from radio to optical 
frequencies using a synchrotron model that entails a 
log-parabolic distribution of electron energies.
A significant correlation among the two fitted spectral parameters ($a$, $b$) of  log-parabolic 
curves and a negative trend among the peak frequency and spectral curvature parameter, $b$, emphasize 
that the SEDs of blazars are fitted well by log-parabolic curves.
On considering each model parameter that could be responsible for changes in the observed 
SEDs of these blazars, we find that changes in the jet Doppler factors are most important.            

\end{abstract}


\section{Introduction}

Studies of the two types of blazars, BL Lac objects and flat spectrum radio quasars (FSRQs), 
have shown that their Spectral Energy Distributions (SEDs) are characterized by a double peaked
luminosity structure \citep[e.g.][]{ghisellini1997}. The BL Lacs and FRSQs classes are defined 
according to the absence or presence of strong broad emission lines in their optical/UV spectrum, 
respectively. The  shapes of these bumps are characterized in the log($\nu F_{\nu}$) vs.\ log   
$\nu$ plot by a smooth spectrum extending through several frequency decades. The low-energy peak 
is well explained by synchrotron emission from relativistic electrons in a jet closely  aligned 
to the line of sight, with bulk Lorentz factor, $\Gamma$ of order of 10--100 
\citep[e.g.,][and references therein]{maraschi1992, ghisellini1993, hovatta2009}.  Moreover, the
position of the low-energy peak leads to a further classification of the
blazars into three categories, depending on the peak frequency of their synchrotron bump, 
$\nu_{peak}^S$: low-synchrotron-peaked (LSP) sources
with $\nu_{peak}^S$ $<$ 10$^{14}$ Hz; intermediate-synchrotron-peaked (ISP)
the sources, those with 10$^{14}$ Hz $<$ $\nu_{peak}^S$ $<$ 10$^{15}$ Hz; and
high-synchrotron-peaked (HSP)  sources, with $\nu_{peak}^S$ $>$ 10$^{15}$
Hz. This scheme is an extension of the classification introduced by \citet{padovani1995}
 for BL Lacs \citep[see][for details]{abdo2010a}.

The high energy (hard X-ray and $\gamma$-ray component  of the SED is 
usually explained as arising from inverse 
Compton scattering  of the same electrons producing the synchrotron emission.
 These electrons interact either with the
synchrotron photons themselves \citep[synchrotron self-Compton, SSC; e.g.,][]{marscher1985}  
or with external photons originating in the local environment
(external Compton, EC). In the latter case soft photons can be produced
directly by the accretion disk \citep[e.g.][]{dermer1993}, or indirectly, for instance those
reprocessed by the broad line region (BLR) \citep[e.g.][]{sikora1994}, or by the dust
torus \citep{blazejowski2000}. Alternative hadronic models, where the $\gamma$-rays 
are produced by high-energy protons, either via proton synchrotron
radiation or via secondary emission from photo-pion and photo-pair-production
reactions, have also been proposed \citep[see][and the references therein for a 
review]{bottcher2007}. 

The  shape of these bumps is characterised in the Log($\nu F_{\nu}$) vs. Log $\nu$ 
plot by a smooth spectrum extending through several frequency decades. Below and above 
the peaks the spectrum can be approximated with simple power law profiles with spectral 
indices above and below the peaks \citep{ghisellini1996}, and such power law spectra are 
naturally produced if the the emitting electrons follow a power law distribution of energy. 
So, the broad band SED of blazars can be well approximated by a simple parabolic function 
with logarithms of its variables \citep[e.g.,][and references therein]{massaro2004, 
massaro2006, tramacere2007}.  The log-parabolic function is one of the simplest ways to 
represent curved spectra and under simple approximations, can be obtained via a statistical 
electron acceleration mechanism where the acceleration probability decreases with the particle energy 
\citep{massaro2006}.   

The emission from blazars is known to be variable at all wavelengths. The flux
variability is often accompanied by spectral changes. These SED changes are
very likely associated with changes in the spectra of
emitting electrons that arise from differences in the physical parameters of the jet. 
Hence modelling of blazar broadband spectra  is required to
understand the extreme conditions within the emission region.  Not only is
the broadband SED crucial to this understanding, but variability information is also needed 
to allow us to describe how high emission states arise and how they
differ from the low states. This type of study is most important in discriminating
between models.  Since it is reasonably to assume that only a few parameters of the model 
will change significantly between different states of the same source, such comparative
modelling of broadband  spectra allows us to put rather tight constraints on those
model parameters that are likely to change, at least under the
assumption that all other parameters are held
fixed for the different model fits \citep[e.g.][]{mukherjee1999, petry2000, hartman2001}.

In this paper, we focus on the study of changes in the  SEDs of BL Lacs and FSRQs 
corresponding to observed changes in their respective optical fluxes. The SED 
variations are expected to be produced by changes in the spectra of the emitting 
electrons which in turn arise from variations in the physical parameters of the  
emission region. Therefore we needed to first construct models and then investigate 
how the SEDs observed in different states might be explained through  changes in 
these parameters. More explicitly, we modelled the observed spectra of blazars from 
radio to optical frequencies using a synchrotron model with the emitting electrons 
following a log-parabolic distribution of energy. This allows us to
try to estimate the factors responsible for short term optical flares or short-term 
variability (STV) in blazars. In this paper, we will strictly concentrate on the 
low energy hump (synchrotron emission) portion of blazar SEDs. A complete investigation 
of the entire broad band SEDs of these objects, including X-ray and $\gamma$-ray 
data, will be performed in the future.

This paper is structured as follows. Section 2 provides a brief description of the 
multi-frequency data we employed. In Section 3, we discuss the SED modelling and Section 4 
provides our results.  Our discussion is given in Section 5 and we present our 
conclusions in Section 6.


\section{Multi-frequency Data }

The multi-frequency SED data of our 10 blazars span  a frequency range between
radio and optical, including mm, sub-mm and infra-red (IR) data.  These 
measurements were all taken between  2008 September and 2009 June. Our sample consists 
of five BL Lacs and five FSRQs,  eight of which are LSPs and two of which are
ISPs. We selected these sources as they were bright
enough for us to obtain  day to day 
flux coverage in B, V, R, and I passbands.  Certainly this number of objects is too small to provide
clearly representative samples of the BL Lac 
and FSRQ classes and there may be unknown biases in our two groups.  Despite the small sample sizes, some general behaviours
might be noted in these data and then investigated in more detail in the future with larger
samples.  The short term variability (STV) in flux as well as in 
colour of these blazars was recently reported by \citet{rani2010b}. We used this
optical data in our SED study and added to them data collected at other frequencies over 
the same time period.  Radio flux densities at 4.5 GHz, 8 GHz and 14.5 GHz frequencies 
are obtained from University of Michigan Radio Astronomy Observatory 
(UMRAO\footnote{http://www.astro.lsa.umich.edu/obs/radiotel/umrao.php}) data base. The mm and 
sub-mm data are provided by the Submillimeter Array (SMA) Observer 
Center\footnote{http://sma1.sma.hawaii.edu/callist/callist.html} 
data base. The near-IR  data are collected from the monitoring provided by the
Small and Moderate Aperture Research Telescope System 
(SMARTS\footnote{http://www.astro.yale.edu/smarts/glast/targets.html}). 

For investigating the impact of the variations in flux on the source parameters
used for modelling the SEDs we always ascertained two SEDs of the same source
that are characterized by a change of at least 0.3 magnitudes in the optical R-band.
The brighter  one is denoted as the high-state and the fainter as the low-state.
The time periods during which the different SEDs of all the sources were obtained
are listed in the first column of 
Table 1 immediately below the name of each of the blazars.  Details of their
optical properties can be found in \citet{rani2010b}. We note that while we tried to obtain and utilize data taken simultaneously,
this was rarely possible, and the temporal intervals over which the data needed
to produce an SED in a low or high state could be found range from three days through
three months.  Clearly, this lack of simultaneity can lead to substantial uncertainties 
in the analysis.

\section{SED Modelling}

We performed the spectral analysis using a homogeneous synchrotron emission model 
with log parabola (LP) energy distribution of emitting electrons to fit the lower 
energy part of observed spectra, i.e., synchrotron spectra. The best fit model 
to each blazar in each of two states was 
obtained by a numerical SSC code \citep{tramacere2007, tramacere2009}. We
used an SED code\footnote{http://tools.asdc.asi.it/} available on-line.

The simplified model assumes that radiation is produced within a single blob in the
jet, which is taken to be moving relativistically at small angle along the line of sight of the
observer. Thus the observed radiation has a Doppler  boosting factor 
$\delta$ = [$\Gamma (1 - \beta ~cos\theta)]^{-1}$, where $\beta$ is the 
 velocity of the source divided by the velocity of light,  $\Gamma$ is the Lorentz
 factor, and 
$\theta$ is the angle between the line of sight of observer and direction 
of motion of the source.   The observed SED of blazars has a double peaked 
structure and a simple analytical function that can model the shape of 
these broad peaks is a parabola in the logarithms of the variables, i.e., a 
log parabola (LP). This function has three spectral parameters and can be 
defined as \citep{massaro2006} 
\begin{equation}
F(E) = K E^{-(a + b~log(E))} ~~ {\rm keV} 
\end{equation}
where $a$ is the photon index at an arbitrary energy (usually taken to be 1 KeV) and $b$ 
is the measure of 
spectral curvature of the observed radiation. One advantage of the LP functional form compared 
to other option is 
that in LP formulation the curvature around the peak is characterized by a single parameter 
$b$, while in other models it is characterized by more complex functions 
\citep[e.g.][]{sohn2003, massaro2006}; however, the limitation is that it is 
more suited for distributions symmetrically decreasing with respect to peak 
frequency, than is the norm for synchrotron spectra.    Although more complex
models, such as a power-law plus log-parabola model, should fit the mm and longer
wavelengths data better, we do not have good coverage in those bands; further, the simpler
log-parabola model has a clear physical motivation, which we now discuss.

\citet{massaro2004} showed that the log-parabolic spectrum is analytically 
related to the statistical acceleration mechanism. It is obtained when the 
acceleration probability, $p$, depends upon energy itself, i.e., 
\begin{equation}
p_{i} = g/\gamma_{i}^{q} ~~~~~ (i = 0, 1, 2, ...) ,
\end{equation}
where $g$, and $q$ are positive constants.  
Such a situation can naturally occur when particles are 
confined by a magnetic field that has a confinement efficiency that decreases with 
an increase in the
gyration radii of the accelerating particles \citep{massaro2006}.  The integrated
energy distribution of such accelerated particles is given by 
\begin{equation}
N(> \gamma) = N(\gamma / \gamma_{0})^{-[s-1 +~ r~log(\gamma / \gamma_{0})]},
\end{equation}
where $\gamma_{0}$ is the minimum Lorentz factor in Eq.\ (2), and $r$ and $s$ are the spectral 
parameters of the electron population and are related to those of the emitted radiation 
as \citep{massaro2006} 
\begin{equation}
a = (s-1)/2 ~~~~~~ 
{\rm and} ~~ ~~~~~~b = r/4 .
\end{equation}



With these approximations, we can completely specify the SSC model with the following  
parameters: magnetic field intensity ($B$), size of emission region ($R$), Doppler boosting factor 
($\delta$), the log-parabola spectral indices ($s$ and $r$), the number density of emitting 
electrons (N), the redshift of the source, $z$, and particle Lorentz factors, $\gamma_{min}$, 
$\gamma_{max}$ and $\gamma_{0}$.   
While fitting the SED of blazars using these parameters, we know the value of $z$ and we
hold $R$, $\gamma_{min}$, and $\gamma_{max}$ fixed.  It is clear that there is a cut-off in 
the low energy part of synchrotron spectra at a frequency around 10$^{11}$ Hz due to synchrotron 
self-absorption \citep[e.g.][and references therein]{ghisellini1999, tavecchio2002}, so we consider 
this self-absorption in modelling the SED of blazars. 
Because we are looking at short term
variability over the course of a just a few months at most, we feel it justified to assume that
$R$ is essentially constant for each source and we find that the values of  $\gamma_{min}$, and $\gamma_{max}$
we used for every case give good ranges for the electron energies needed to fit any of the SEDs and so
they are fixed as well.   The values of $r$, $s$ and $\gamma_{0}$ are rather 
tightly constrained to fit the observed slopes 
and we then made over than 30 different models
for each source to produce the values of all the parameters that yield a ``best fit''.

A direct observation of the synchrotron peak  will give a 
significant help to constrain the model parameters, but unfortunately,  observations of blazars 
in infrared bands, above all, in the  medium and far infrared, are not so common as are the radio,
optical and near IR ones.  Occasionally this gap can be covered by space-based telescopes such as 
Herschel \citep{pilbratt2010}.   In addition, for FSRQs the SEDs are sometimes 
overwhelmed by thermal emission from an accretion disk, particularly during low-states \citep{malkan1982}, but our code is not able to consider this contribution while modelling the SEDs of blazars.  This additional
contribution may explain the excess of optical/UV emission with respect to the best-fit model for
some FSRQs.  
The values of all these  parameters that provide good fits for the two different SEDs of 
all of the  eight LSPs and two ISPs in our sample are listed in  Table 1. The 
fitted SEDs are shown in Figs.\ 1 $-$ 3.

\begin{table*}
\caption{ Parameters of SED models using log-parabola fits }
\scriptsize
\noindent
\begin{tabular}{llcccclccccccc} \hline
Source & $ <R_{mag}>$  &Log R& B& $\delta$ & N&  z  &   Log & Log & Log   &    r  &    s  & Log $\nu$F$_{\nu}$$_{Peak}$&Log $\nu_{Peak}$      \\
                       &         & (cm)   & (Gauss) &  & (cm$^{-3}$)&     & $\gamma$$_{min}$ & $\gamma$$_{max}$ & $\gamma$$_{0}$  &       &  & (erg cm$^{-2}$ s$^{-1}$) & (Hz)        \\\hline
3C 66A            &        &       &        &        &      &         &      &    &      &       &      &         &       \\
20-30 Oct. 2008   & 14.60  &  17.0 &  0.10  &  15.2  &  30  &  0.444  & 0.2  & 6  & 4.00 &  0.5  & 3.00 &  -10.71 & 14.66  \\
20-22 Jan. 2009   & 14.05  &  16.8 &  0.11  &  12.2  &  30  &  0.444  & 0.2  & 6  & 4.10 &  0.6  & 3.00 &  -10.40 & 14.80  \\
AO 0235$+$164     &        &       &        &        &      &         &      &    &      &       &      &         &         \\
Sept. 2008        & 16.00  &  17.2 &  0.09  &  20.0  &  40  &  0.94   & 0.2  & 6  & 3.38 &  1.9  & 3.65 &  -10.21 & 13.07    \\
20-30 Oct. 2008   & 15.00  &  17.2 &  0.09  &  21.0  &  40  &  0.94   & 0.2  & 6  & 3.42 &  1.9  & 3.65 &  -10.07 & 12.95    \\
20-22 Jan. 2009   & 17.00  &  17.2 &  0.08  &  18.0  &  35  &  0.94   & 0.2  & 6  & 3.34 &  1.9  & 3.40 &  -10.58 & 13.03    \\
PKS 0420$-$014    &        &       &        &        &      &         &      &    &      &       &      &         &           \\
20-30 Oct. 2008   & 16.85  &  17.0 &  0.60  &  18.0  &  35  &  0.915  & 0.2  & 6  & 2.80 &  1.1  & 3.10 &  -10.81 & 12.93     \\
Jan. 2009         & 16.45  &  17.0 &  0.60  &  20.0  &  37  &  0.915  & 0.2  & 6  & 2.80 &  1.1  & 3.20 &  -10.69 & 12.95     \\
S5 0716$+$714     &        &       &        &        &      &         &      &    &      &       &      &         &           \\
20-30 Oct. 2008   & 15.30  & 17.0  & 0.09   &   14.0 &   20 & 0.31    & 0.2  & 6  & 3.60 &  0.9  &  3.20&  -10.78 &  13.60    \\
Mar.-May 2009     & 13.60  &  17.0 & 0.085  &   15.0 &   15 & 0.31    & 0.2  & 6  & 4.12 &  0.75 &  3.34&  -10.29 &  14.55    \\
PKS 0735$+$178    &        &       &        &        &      &         &      &    &      &       &      &         &            \\
May 2009          & 15.90  &  17.13&  0.07  &  12.0  &  20  &  0.424  & 0.2  & 6  & 4.00 &  0.5  & 3.00 &  -11.20 & 14.41      \\
June 2009         & 15.60  &  17.13&  0.07  &  12.7  &  20  &  0.424  & 0.2  & 6  & 4.00 &  0.5  & 3.02 &  -11.14 & 14.52      \\
OJ 287            &        &       &        &        &      &         &      &    &      &       &      &         &            \\
20-30 Oct. 2008   & 14.20  &  17.0 &  0.10  &  18.0  &  20  &  0.306  & 0.2  & 6  & 3.40 &  1.5  & 3.25 &  -10.07 & 13.51      \\
Mar.-Apr. 2009    & 14.70  &  17.0 &  0.12  &  19.0  &  25  &  0.306  & 0.2  & 6  & 3.30 &  1.5  & 3.45 &  -10.06 & 13.21      \\
4C 29.45          &        &       &        &        &      &         &      &    &      &       &      &         &            \\
20-29 Mar. 2009   & 17.00  &  17.0 &  0.50  &  15.0  &  30  &  0.729  & 0.2  & 6  & 2.80 &  1.3  & 2.95 &  -10.77 & 12.96       \\
May 2009          & 18.20  &  17.0 &  0.45  &  15.0  &  25  &  0.729  & 0.2  & 6  & 2.80 &  1.1  & 2.95 &  -11.16 & 12.91      \\
3C 279            &        &       &        &        &      &         &      &    &      &       &      &         &            \\
Apr. 2009         & 15.40  &  17.0 &  0.60  &  17.0  &  30  &  0.5362 & 0.2  & 6  & 2.70 &  1.7  & 3.20 &  -10.20 & 12.82      \\
June 2009         & 16.00  &  17.0 &  0.65  &  17.0  &  30  &  0.5362 & 0.2  & 6  & 2.70 &  1.4  & 3.30 &  -10.43 & 12.53      \\
PKS 1510$-$089    &        &       &        &        &      &         &      &    &      &       &      &         &             \\
Apr. 2009         & 15.40  & 16.8  & 0.50   & 14.0   & 30   & 0.36    & 0.2  & 6  & 3.10 &  1.0  & 3.155&  -10.69 & 13.41      \\
June 2009         & 16.00  & 16.8  & 0.50   & 14.0   & 30   & 0.36    & 0.2  & 6  & 3.10 &  0.8  & 3.155&  -11.02 & 13.45      \\
3C 454.3          &        &       &        &        &      &         &      &    &      &       &      &         &            \\
4-10 Sept. 2008   & 14.70  &17.2   &0.50    &23.0    &30    &0.859    &0.2   &6   &2.80  & 2.3   & 3.35 &   -9.46 &  13.22    \\
20-30 Oct. 2008   & 15.40  &17.2   &0.48    &21.0    &30    &0.859    &0.2   &6   &2.70  & 1.9   & 3.00 &   -9.81 &  12.75     \\\hline              
\end{tabular} \\
Log R(cm) : Size of emitting region \\
B(Gauss)  : Magnetic field        \\
$\delta$  : Doppler boosting factor   \\
N(cm$^{-3}$) : Number density of emitting electrons \\
z : redshift \\
$\gamma$$_{min}$, $\gamma$$_{max}$ : Minimum and maximum values of Lorentz factor \\
$\gamma$$_{0}$ : the Lorentz factor corresponding to the energy where s is evaluated \\
r and s : spectral parameters of the electron population (see text) \\
$\nu_{Peak}$ : Synchrotron peak frequency {\bf in the rest frame of source}

\end{table*}

\section{Results}

We fit the radio to optical SEDs of 10 blazars with
a synchrotron model. The SED curves of all the sources are shown in 
Fig. 1 $-$ 3 and the fitted parameters are listed in Table 1.  Detailed 
multiband optical STV studies of the fluxes and colours 
of all of these blazars over the same time period is reported in \citet{rani2010b}. 
There we showed that the colour versus brightness correlations support the hypothesis 
that these BL Lacs tend to be bluer with an increase in brightness, while these FSRQs show 
the opposite trend.  Our intraday variability (IDV) study of these sources in the optical R-band 
have been recently reported in \citet{rani2011}.
All of these  blazars belong to the First Fermi LAT
  catalogue \citep[1FGL Catalog,][]{abdo2010a}, and their 
spectral properties (hardness, curvature and variability) at GeV energies, 
have been established by \citet{abdo2010b}. The distribution of photon indices 
($\Gamma$) above 100 MeV is found to correlate strongly with blazar subclass.
Also, the spectral indices  tend to be harder when the flux is 
brighter for both FSRQs and BL Lacs. 

We now summarize some previous observations of each of these sources.
Since the optical flux, colour and spectral variability 
of these  LSPs and ISPs  have already been discussed in \citet{rani2010b}, and IDV studies 
are in 
\citet{rani2011}, here we highlight some results from earlier studies 
of these sources at other wavebands. \\
  
\noindent 
{\bf 3C 66A } \\
This ISP blazar is classified as a BL Lac object. The redshift of 3C 66A was reported 
as z=0.444 \citep{miller1978}, but this value remains uncertain
\citep[see][]{bramel2005, yan2010}. 3C 66A is observed in radio, IR, optical, X-rays, and
gamma rays and shows strong luminosity variations \citep[][and references therein]{terasranta2004, 
bottcher2009, rani2010b}. \citet{bottcher2005} organized an intensive multiwavelength campaign 
of the source from 2003 July through 2004 April to understand its broadband spectral behaviour. 
\citet{joshi2007} used a leptonic SSC jet model to reproduce the broadband SED and the observed 
optical spectral variability patterns of 3C 66A and made predictions regarding observable X-ray
spectral variability patterns and $\gamma$-ray emission.   \\

\noindent 
{\bf AO 0235$+$164} \\
The blazar AO 0235+164 at z = 0.94 \citep{nilsson1996} was classified as a BL Lac object
by Spinrad \& Smith (1975). An apparent quasi-periodic oscillation of $\sim$17 days has been 
reported in the X-ray light curves of the source by \citet{rani2009}. The XMM-{\it Newton} 
observations suggested the presence of an extra emission component in the source SED, in 
addition to the synchrotron and inverse-Compton ones. The origin of this component, peaking 
in the UV/soft X-ray frequency range, is probably thermal emission from an accretion disc, 
or synchrotron emission from an inner jet region \citep[][and references therein]{raiteri2006, 
raiteri2008}. \citet{thorn2008} reported a significant correlation between the flux density 
and degree of polarization at optical frequencies and found that at the maximum degree of polarization 
the electric vector tends to align with the parsec-scale radio jet's direction.    \\

\noindent  
{\bf PKS 0420$-$014} \\
PKS 0420-014 is a FSRQ at a redshift z = 0.915 \citep{kuehr1981} that has been
observed in optical bands since 1969. There is a series of papers reporting the optically active
and bright phases of source and semi-regular major flaring cycles \citep[e.g.][and references 
therein]{webb1988, villata1997, raiteri1998}. \citet{rani2010b} reported that the source 
tend to be redder with increase in its
 brightness. Recently rapid optical and radio brightening of the blazar has been 
reported by \citet{bach2010}.  \\

   \begin{figure*}
   \centering
\epsfig{figure = 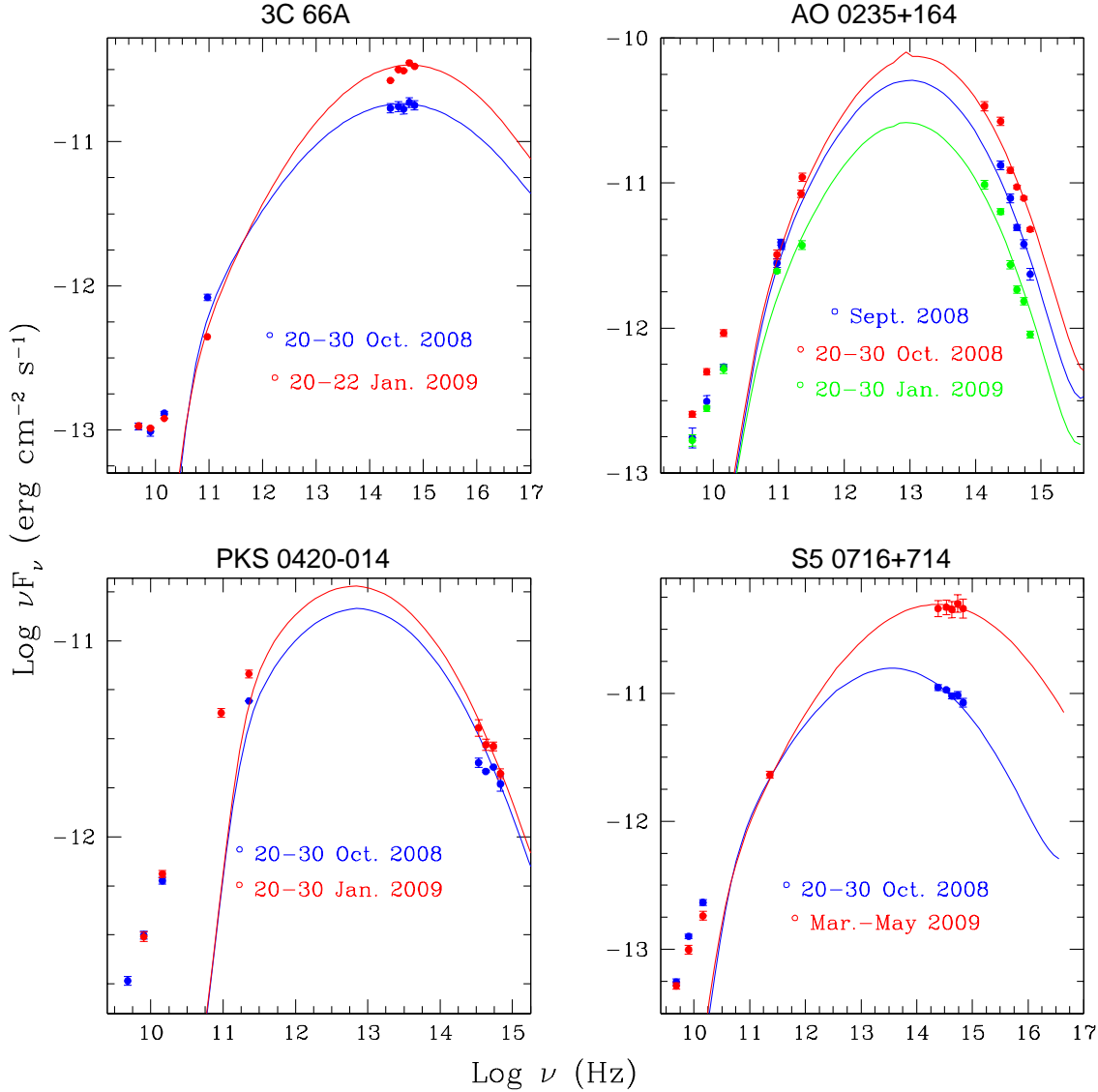,height=16.cm,width=16.cm,angle=0}
   \caption{The modelled SED curves of the blazars 3C 66A, AO 0235$+$164, PKS 0420$-$014 and 
S5 0716$+$716. The points represent the observed data while the best fitted model is shown 
by line.  Since the low frequency part ($\leq 10^{11}$ Hz) of blazars SEDs is governed by 
synchrotron-self absorption mechanism, so the modelled SED is steeper below a frequency of 
10$^{11}$ Hz (see text for details). 
            }
    \end{figure*}

\noindent 
{\bf S5 0716$+$714} \\
S5 0716+714 was identified as an ISP \citep{giommi1999}, 
at z = 0.31 $\pm$ 0.08 \citep{nilsson2008}. 
However, Nieppola et al.\ (2006) studied the SED distribution of a large sample of BL Lac objects 
and categorized 0716$+$714 as a LSP. Two strong $\gamma$-ray flares on September and October 2007 from 
the source have been detected by AGILE  \citep{giommi2008, chen2008}. They have also 
carried out the SED modelling 
of source with two synchrotron self-Compton (SSC) emission models, representative of a slowly 
and a rapidly variable component, respectively. \citet{villata2008} reported optical-IR SED study of the 
source during the GASP-WEBT-AGILE campaign in 2007. \citet{gupta2009} found high probabilities of 
quasi-periodic components of $\sim$25$-$73 minutes in the optical light curves observed by 
\citet{montagni2006}.  Recently, nearly periodic oscillations of $\sim$15 minutes have been reported in 
the optical light curve of the source by \citet{rani2010a}.  \\

   \begin{figure*}
   \centering
\epsfig{figure = 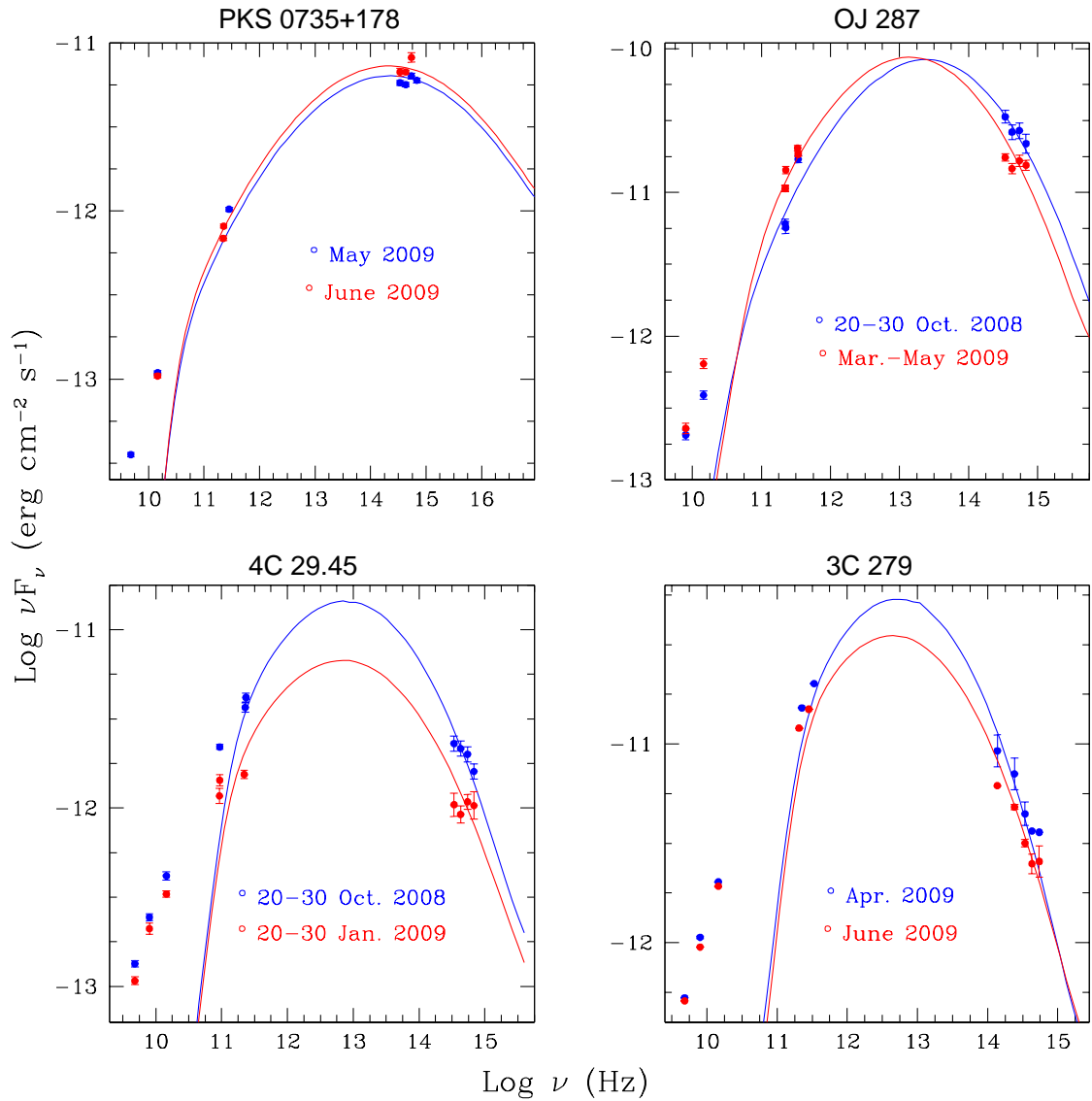,height=16.cm,width=16.cm,angle=0}
   \caption{The modelled SED curves of the blazars PKS 0735$+$178, OJ 287, 4C 29.45 and 3C 279.
              }
    \end{figure*}

\noindent 
{\bf PKS 0735$+$178} \\
PKS 0735$+$718 is a highly variable quasar and belongs to the category of BL Lac objects 
\citep{carswell1974}. There have been several papers concerning its redshift 
determination \citep[e.g.][and the references therein]{carswell1974, falomo2000} 
with the most recent result of $z = 0.424$ for PKS 0735$+$718 found using a Hubble Space 
Telescope (HST) snapshot image \citep{sbarufatti2005};  using that redshift, we classify
this source as an ISP. 
The synchrotron part of SED of the source peaks at  IR-UV, while the low X-ray variability with 
respect to the high optical-IR variations, supports the idea that X-rays are produced
by an inverse Compton mechanism \citep{bregman1984, madejski1988}.
Using  ten year long multiband optical data of this source \citet{ciprini2007} studied the 
 multicolour behaviour, spectral index variations and their correlations with  brightness. \\

\noindent 
{\bf OJ 287} \\
OJ 287 (0851$+$202) (z = 0.306) is among the most extensively observed and best studied BL
Lac objects with respect to variability.  As it is very bright, it is among the very few AGN's whose optical observations are available for more than a century \citep{sillanpaa1996, gupta2008}. A binary black hole 
model has been proposed to explain the $\sim$12 years  optical periodicity \citep{sillanpaa1996, valtonen2008}.  
An intense optical, infrared, and radio variability study of the source for a time period between 
1993 to 1998 was been reported in \citet{pursimo2000}. They detected two major optical outbursts in 
November 1994 and December 1995 even though the radio flux was very low during the period of outbursts. 
Using the multi-waveband flux and linear polarization observations and combining 
with sub-mm polarimetric images,  \citet{agudo2010} argued that the location of the $\gamma$-ray 
emission in prominent flares is $>$ 14 pc from the central black hole.   \\

\noindent 
{\bf 4C 29.45} \\
4C 29.45 \citep[z = 0.729;][]{wills1983}, belongs to the category of FSRQs. 
Short and long term variability at both IR and optical bands has been 
observed in this source \citep{noble1996, ghosh2000}.  Its optical flux and multi-colour variations 
were studied by \citet{fan2006}. They reported amplitude
variations of $\sim$4.5$-$6 mag in all passbands (U, B, V, R, I) and also
suggested that there were possible periods of 3.55 and 1.58 yr in the long-term optical 
LC of the source. The short term multi-band 
optical colour-brightness studies showed that the source tend to be redder with 
increase in its brightness \citep{rani2010b}. Very recently, a GeV flare has been detected in this blazar
\citep{ciprini2010}.   \\

\noindent 
{\bf 3C 279} \\
3C 279 is both one of the most extensively observed and most violent variable FSRQs across the whole
electromagnetic spectrum; it is at a redshift of $z = 0.536$. A comparison of radio to Gamma-ray 
SEDs of the source during low and high states of activity in the WEBT campaign of 2006 is presented 
in \citet{bottcher2007}. \citet{collmar2010} carried out multifrequency variability study of 
source during January 2006. A complete compilation of all simultaneous SEDs of 3C 279 collected 
during the lifetime of CGRO, including modelling of those SEDs with a leptonic jet model, is 
presented in \citet{hartman2001}. Recently, \citet{abdo2010c} report the coincidence of a 
$\gamma$-ray flare with a dramatic change of optical polarization angle. This provides evidence 
for co-spatial of optical and $\gamma$-ray emission regions and indicates a highly ordered jet 
magnetic field.  \\

\noindent
{\bf PKS 1510$-$089} \\
PKS 1510$-$089 is classified as a FSRQ at a redshift of z = 0.361 \citep{thompson1990}. 
It also belongs to the category of highly polarized quasars and already detected  in the
MeV--GeV energy band by the EGRET instrument on board the Compton Gamma-Ray Observatory \citep{hartman1992}. 
The inverse Compton component dominated by the $\gamma$-ray emission, and the synchrotron 
emission peaked around IR frequencies, even if there is clearly visible in this source 
a pronounced UV bump, possibly caused by the thermal emission from the accretion disk 
\citep{malkan1986, pian1993}. The multi-wavelength flux and SED variability study of the 
source during its high $\gamma$-ray activity period between 2008 September and 2009 June 
has been  reported by \citet{abdo2010c}.       
Recently, \citet{dammando2011} reported the detailed analysis of multi-frequency coverage of 
extreme gamma-ray activity from the source observed by AGILE in March 2009. They have also
carried out broad band SED study of the source and found thermal features in the optical/UV 
spectrum of the source during high $\gamma$-ray state.   \\  

\noindent 
{\bf 3C 454.3}  \\
3C 454.3 is a FSRQ at a redshift of $z = 0.859$, and is one of the most intense and variable of these
sources \citep[][and references therein]{pian2006, villata2007, gupta2008}. 
It is the brightest and most variable blazar at GeV energies \citep[e.g.][]{vercellone2008,
raiteri2008, vercellone2010}.  The multi-flare variability and SED study of the source in 
December 2007 and December 2009 is reported by \citet{donnarumma2009} and \citet{pacciani2010}, 
respectively. The flaring behaviour of the source across the whole electromagnetic spectrum is 
studied by \citet{jorstad2010}. They conclude that the emergence of a superluminal knot from the 
core yields a series of optical and high-energy outbursts, and that the millimetre-wave core lies 
at the end of the jet’s acceleration and collimation zone.

   \begin{figure}
   \centering
\epsfig{figure = 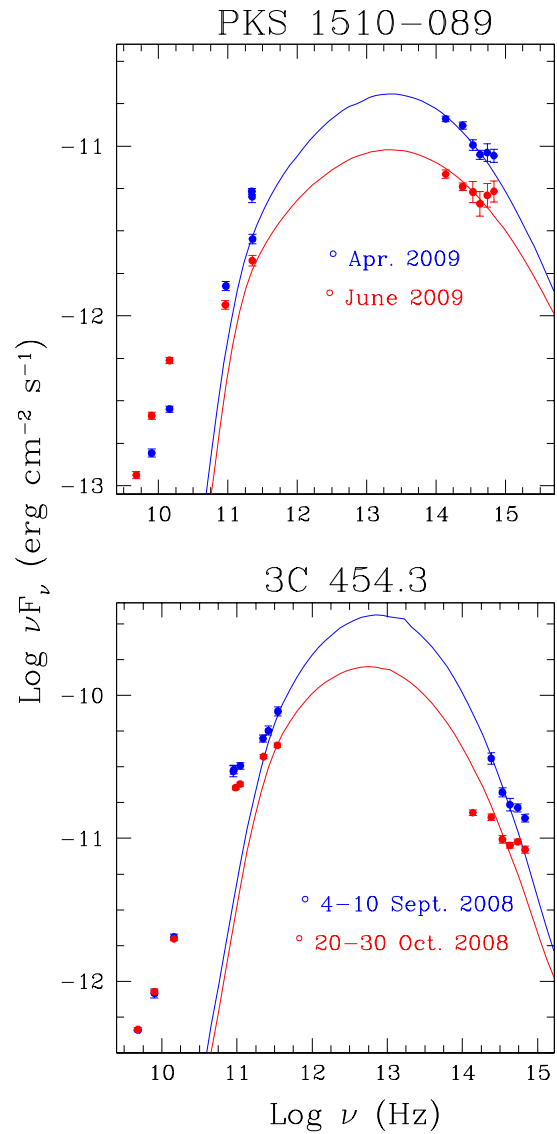,height=16.0cm,width=16.0cm,angle=0}
   \caption{Modelled SED curves of the blazars PKS 1510$-$089 and 3C 454.3.  The last 
three points in all of these SEDs are almost certainly dominated by contribution from the disk emission, so we exclude these points 
while modelling  the SEDs of these two FSRQs. 
              }
    \end{figure}

\section{Discussion}

\subsection{Statistical particle acceleration and log-parabolic spectra}
\citet{massaro2004} showed that when the acceleration efficiency is inversely proportional to 
the accelerating particle's energy itself, then the energy distribution function, i.e., spectrum, 
approaches a log-parabolic 
shape. So, the log-parabolic spectra are naturally produced when the probability 
of statistical acceleration is energy dependent. According to this model, the curvature of emitting electron 
population ($r$) is related to fractional acceleration gain ($\epsilon$) as r $\propto$ 
1/$\epsilon$. Also, E$_{P}$ $\propto$ $\epsilon$ follows a negative trend between E$_{P}$ 
and $b$ \citep[see][]{tramacere2009}, where E$_{P}$ is the peak energy of observed SEDs. 

So, if the above analysis is correct, then one should expect a negative trend between E$_{P}$ or 
$\nu_{P}$ and $b$, where $\nu_{P}$ is the peak frequency of observed SEDs. We analysed the 
correlation between the two quantities using our data and 
found a significant negative correlation between $\nu_{P}$ and $b$ with $r_P = -0.74$ and 
$p$-value = 0.00014, where $r_{P}$ is the linear Pearson correlation coefficient and the $p$-value 
for the 
null hypothesis of no correlation corresponds to a significance level $>$ 99.98$\%$ (Fig.\ 4). So our 
result confirms that the spectral curvature parameter decreases as E$_{P}$ moves towards 
higher energies and means that the blazars' spectra in our cases are fit well 
by the log-parabolic model.      

On the other hand, such a connection of log-parabolic spectra with acceleration may be also 
understood in the framework provided by the Fokker-Planck equation with momentum diffusion 
term \citep{kardashev1962, massaro2006, tramacere2009}. They showed that 
the log-parabolic spectrum results from a Fokker-Planck equation  with a momentum diffusion
term and a mono-energetic or quasi-monoenergetic injection where the diffusion term acts to 
broaden  the  shape  of the peak of the distribution. \citet{kardashev1962} showed that the 
curvature term, $r \propto 1/(Dt)$ where $D$ is the Diffusion term. This relation leads 
to the following trend \citep{tramacere2009} : 
\begin{equation}
ln(E_{P}) = 2~ln(\gamma_{P}) + \frac{3}{5b}
\end{equation}
So the both the momentum-diffusion term, $D$, and the fractional acceleration gain
term, $\epsilon$,
can explain the anti-correlation between $E_{P}$ and $b$.

An interesting check if the log-parabolic curve is actually related to the statistical 
acceleration is the existence of a linear relation between the two spectral parameters 
$a$ and $b$ \citep{massaro2004}. So we checked the correlation among these two quantities 
and found that they are significantly correlated, with 
$r_P = 0.61$ and $p$-value = 0.003 (see Fig.\ 5). So, we consider that in our case the log-parabola 
type spectra are very likely to be  characterized by a full statistical acceleration mechanism working on
the emitting electrons.

 \begin{figure}
   \centering
\epsfig{figure = 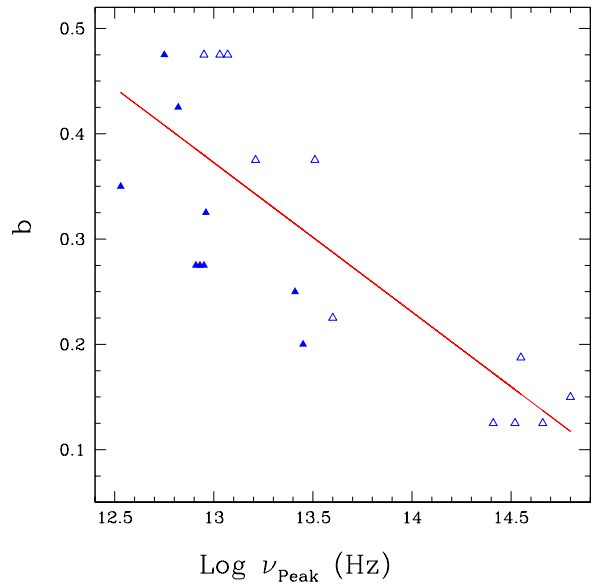,height=8.0cm,width=8.0cm,angle=0}
   \caption{ Curvature ($b$) vs peak frequency ($\nu_{P}$). A clear relation 
can be seen  between the two quantities which is confirmed by a correlation analysis. 
The open symbols in this and subsequent figures represent BL Lacs while the filled symbols stand for FSRQs.
 The straight 
line is the best linear fit with a slope, $m = -0.14$ and constant, $c = 2.22$ ($y  = m x + c$). }
    \end{figure}

 \begin{figure}
   \centering
\epsfig{figure = 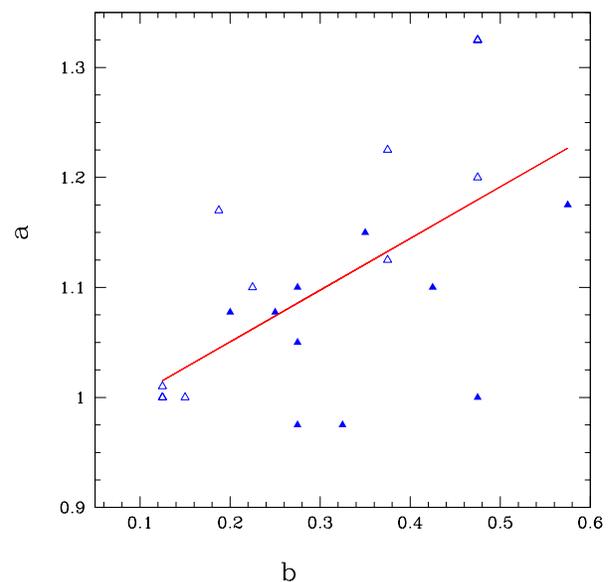,height=8.0cm,width=8.0cm,angle=0}
   \caption{ The  spectral parameters, $a$ and $b$. A clear relation can be seen    
between the two quantities which is confirmed by a correlation analysis. The straight 
line is the best linear fit with $m = 0.47$ and $c = 0.96$.         }
    \end{figure}

   \begin{figure}
   \centering
\epsfig{figure = 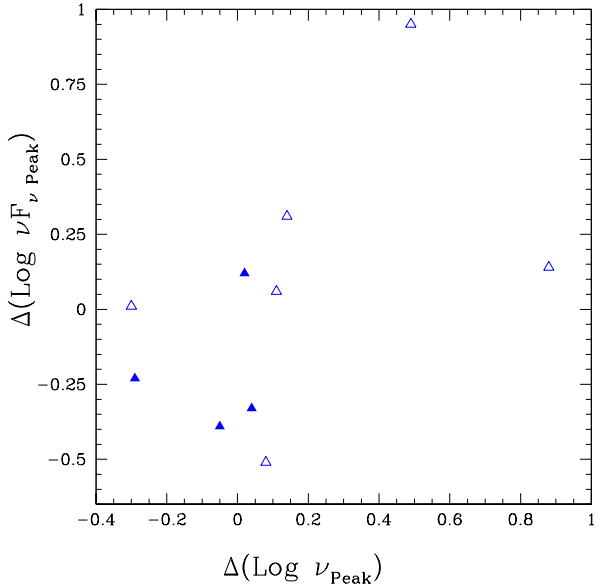,height=8.0cm,width=8.0cm,angle=0}
   \caption{Changes in the peak intensity vs.\ change in peak frequency for 
all blazars (both BL Lacs and FSRQs together).  There is no significant correlation found
between the two quantities.} 
    \end{figure}

\subsection{The cause of blazar SED changes}
There are several possibilities that could be invoked to explain the 
changes in the synchrotron spectra. In the following sub-sections, we 
will discuss some of them in detail. \\

\subsubsection{Evolution of the particle energy density distribution} 

The energy loss of the emitting particles is known to follow $\frac{d\gamma}{dt}$ $\cong$ $-$C$\gamma^{2}$ 
where $\gamma$ is the Lorentz factor and $C$ is a constant for both the synchrotron and IC 
emission \citep{rybicki1979}. The evolution of the 
relativistic particles can be described by energy-dependent Fokker-Planck equation 
\citep[see for instance][p.52]{kembhavi1999} :

\begin{equation}
\frac{\partial (\gamma,t)}{\partial t} = - \frac{\partial}{\partial \gamma} \Bigg( \frac{d \gamma}{dt} n(\gamma,t) \Bigg) + Q(\gamma,t)
\end{equation}
where n($\gamma$,t) is the electron density distribution (a function of energy and time) and 
Q($\gamma$,t) is the time-dependent injection rate. The solutions to the equation above are 
typically rather complex and can be obtained in terms of Green's functions for the 
general case. 


   \begin{figure}
   \centering
\epsfig{figure = 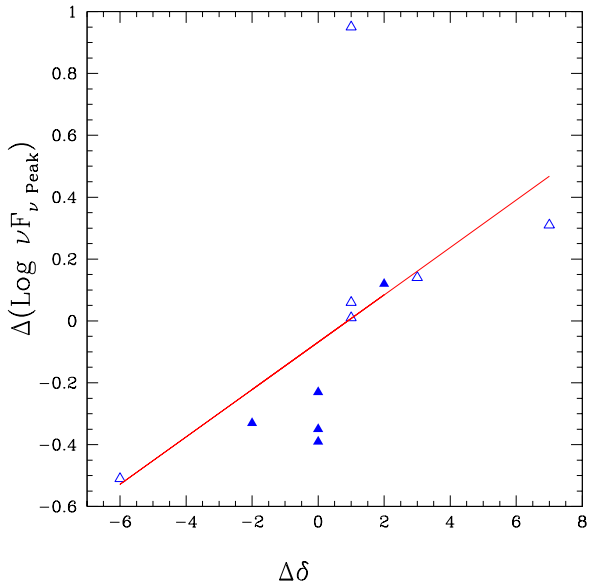,height=8.0cm,width=8.0cm,angle=0}
   \caption{Changes in the peak intensity vs.\ change in Doppler factor for all 
blazars that have multiple SEDs in our sample. There is a clear relation seen between the 
two quantities.
The straight  line is the best linear fit with $m = 0.07$ and $c = -0.07$.   } 
    \end{figure}

   \begin{figure}
   \centering
\epsfig{figure = 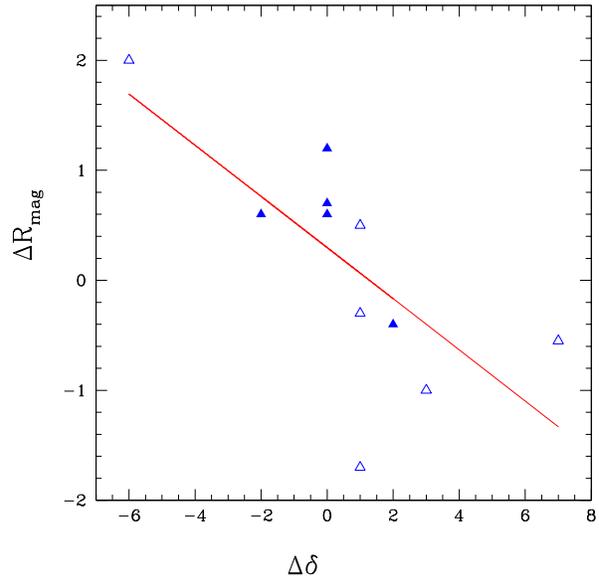,height=8.0cm,width=8.0cm,angle=0}
   \caption{ Changes in the R band magnitude vs.\ change in Doppler factor  
for all blazars that have multiple SEDs in our sample. There is a significant correlation 
between the two quantities, with $r_P = -0.70$ and $p$-value = 0.01.   The straight
line is the best linear fit with $m = -0.23$ and $c = 0.29$. } 
    \end{figure}

If the changes in the blazar SEDs are due to a gradual change of the 
electron energy density distribution due to the synchrotron and IC losses, with no other 
injections meanwhile, then one should see both a total synchrotron energy decrease and 
mostly a decrease in $\gamma_{b}$ of the electron energy distribution (and respectively $-$ 
the frequency of the synchrotron peak emission). On the other hand, freshly injected electrons 
between the observational epochs (their energy should be higher than the average as it is expected 
to decrease in time) should reflect in the opposite behaviour (higher synchrotron emission 
and a peak blue-shift). So, in general one should expect to see a relation between the peak 
intensity and the peak frequency changes. To search for such a relation we used correlation analysis 
statistics (see Fig. 6). The correlation statistics revealed that there is no significant 
correlation between the two quantities as $r_P = 0.54$ corresponding to a   
$p$-value = 0.084,  where $r_P$ is the linear Pearson correlation coefficient and $p$ is a null hypothesis
probability value. As there is no significant relation found between the two quantities, 
 perhaps electron energy density evolution/electron injection is not the primary 
driver of the SED changes.

 \begin{figure}
   \centering
\epsfig{figure = 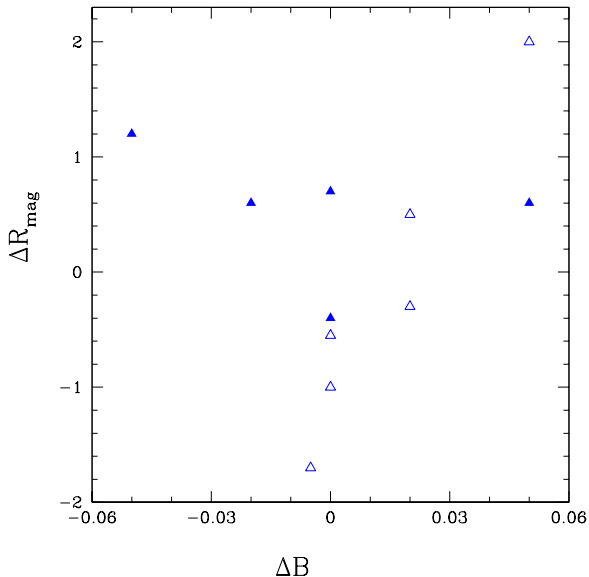,height=8.0cm,width=8.0cm,angle=0}
   \caption{ Changes in the R band magnitude vs.\ change in magnetic field
for all blazars that have multiple SED’s in our sample. There is no significant correlation
between the two quantities with $r_P = 0.11$ and $p$-value = 0.75.}
    \end{figure}

\subsubsection{Change of the Doppler boosting factor}
Another effect that can modify the blazar synchrotron emission is the change of the
Doppler boosting factor, presumably due to a change either the bulk velocity or
 the viewing angle. \citet{raiteri2010} also
found that only geometrical (Doppler factor) changes are capable of explaining
SED variations between two epochs for BL Lac.  The changes
of the synchrotron peak intensity vs.\ Doppler factor variations for our sample of
blazars are shown in 
Fig.\ 7. We can see a clear relation among the two quantities and it is confirmed 
by values of $r_P = 0.51$ and 
$p$-value = 0.05. We also searched for a correlation between change in R band magnitude vs.\ change 
in respective Doppler factor values.  This would be relevant if $<$R$>$ is a somewhat more representative quantity for the synchrotron emission rather than peak intensity. Again, we found significant 
correlation among $\Delta$$R_{mag}$ and $\Delta$$\delta$ (see Fig.\ 8). So we consider 
that Doppler factor changes are a strong driver that can be responsible for 
the SED changes.

\subsubsection{Change of the Magnetic Field}
Since the value of the magnetic field, $B$ is related to the location of the dissipation 
region, if the two SEDs arise from  periods of different activity it is quite possible 
that the emission was produced by blob that dissipates at different location in the jet. 
In that case we expect to find different $B$ values for the different SEDs of our sources. The 
change in $B$ values can also lead to variations in the flux of these sources which will 
be reflected in the two different SEDs of the sources. So we search for a correlation 
between the change in $<$R$>$ and that in $B$ (Fig.\ 9) but we do not find any 
significant correlation among them {for all of out blazars. 
However, we notice an apparent significant correlation among these two parameters for 
the BL Lacs alone, which is 
confirmed by correlation analysis ($r_P = 0.96$, $p = 0.002$ for BL Lacs). 
So we conclude that changes in magnetic field 
strength  may be responsible for the SED changes in the case of BL Lacs}.  

\section{Conclusions}
We have carried out the radio to optical through mm, sub-mm and IR SED studies 
of a sample of ten blazars including five BL Lacs and five FSRQs, eight of which are LSPs
and the other two ISPs. We modelled the 
SEDs of blazars using synchrotron spectra with  log-parabolic distributions. 
We found a significant negative correlation among $\nu_{Peak}$ and the curvature 
term $r$, which implies that the acceleration efficiency of emitting electrons 
is inversely proportional to energy itself; however this correlation can also be 
explained by  the momentum diffusion term in the solution of Fokker-Planck equation.
Also, a significant correlation between the two spectral parameters $a$ and $b$, 
implies that the log-parabolic curve is likely to be related to the statistical acceleration 
of emitting electrons. 

Of course our modelling has significant limitations.  We are assuming that only
one-zone dominates the emission at any given time, and this is unlikely to be
an excellent approximation.  Although we strived for data obtained simultaneously,
this was often impossible to obtain; therefore any variations within the ``high" and
``low" states have been averaged over time bins ranging from days to two months.
This lack of simultaneity could have vitiated our results but only appears to have
added scatter to all of the Figures 4--9.
 
We considered each likely factor that could be responsible for the changes in 
observed SEDs of blazars. If the electron energy density evolution governs the 
SED changes then one should expect a correlation between change in peak intensity 
vs change in peak frequency. Since we do not observe any such correlation
we consider that probably the evolution of electron energy density cannot be 
responsible for the observed SED changes. Also for our entire sample of 
blazars, changes in $<$R$_{mag}$$>$ are not correlated with the respective 
changes in $B$, so the change in magnetic field strength is probably not 
responsible for the SED changes we saw.  However, for the BL Lacs alone the SED changes 
may be driven by changes in $B$. 
We find that the change in Doppler factor is significantly correlated with the change in 
peak intensity as well as with $<$R$_{mag}$$>$. So it is reasonable to suggest
that the change in Doppler 
factor (either bulk velocity or viewing angle) is the primary driver that governs the SED changes in 
short term  variability (STV) in blazars.

\section*{Acknowledgments} 

We thank the referee, Dr.\ Paolo Giommi, for several helpful suggestions.
BR is thankful to Prof.\ D.\ C.\ Srivastava for his valuable suggestions and encouragements and 
to Mr.\ Ravi Joshi for help while finalizing the text.
This research was partially supported by Scientific Research Fund of the 
Bulgarian Ministry of Education and Sciences (BIn - 13/09, DO 02-85 and DO02-340/08)
and by Indo--Bulgaria bilateral scientific exchange project INT/Bulgaria/B$-$5/08 funded
by DST, India. RB and AS acknowledge the kind hospitality of ARIES, Nainital, India.
This research has made use of data from the University of Michigan Radio Astronomy Observatory 
which has been supported by the University of Michigan and by a series of grants from the National 
Science Foundation, most recently AST-0607523. BR is very grateful to Margo Aller for providing the 
data at radio frequencies.  
The Submillimeter Array is a joint project between the Smithsonian Astrophysical Observatory and 
the Academia Sinica Institute of Astronomy and Astrophysics and is funded by the Smithsonian 
Institution and the Academia Sinica. We used this data in our research. 
The Australia Telescope Compact Array is part of the Australia Telescope which is funded by the 
Commonwealth of Australia for operation as a National Facility managed by CSIRO. The efforts of 
ATNF staff in maintaining the ATCA calibrator database (http://www.narrabri.atnf.csiro.au/calibrators/) 
are gratefully acknowledged.
SMARTS data are made available by Yale University at http://www.astro.yale.edu/smarts/fermi through 
Fermi GI grant 011283.


\label{lastpage}

\end{document}